\documentclass[prd,tightenlines,floatfix,showpacs,preprintnumbers,nofootinbib,eqsecnum]{revtex4}

  \usepackage{amssymb}
   \usepackage{amsmath} 
    \usepackage{amsfonts}
    \usepackage{epsfig}
     \usepackage{color}
      \usepackage{bm} 
        \usepackage{xcolor}
        
\usepackage{booktabs}
\usepackage{multirow}
\usepackage{slashed}
\usepackage{comment}

\usepackage{cancel}
\usepackage[normalem]{ulem}

\usepackage{amscd}
\usepackage{graphicx}
\usepackage{subfigure}
\usepackage{hyperref}

\def\doublespace{\def\baselinestretch{1.6}\large\normalsize}
\def\normalspace{\def\baselinestretch{1.0}\normalsize}

\def\Caption#1{
  \normalspace
  \vskip-1mm\caption{\sl#1}\vskip-1mm
  \doublespace
}

\def\BA{\begin{eqnarray}}
\def\BE{\begin{equation}}
\def\BF{\begin{figure}[htb]}
\def\BT{\begin{table}[htb]}
\def\EA{\end{eqnarray}}
\def\EE{\end{equation}}
\def\EF{\end{figure}}
\def\ET{\end{table}}

\def\la{\langle}
\def\ra{\rangle}
\def\fm{\,\mbox{fm}}

\def\mb{\,\mbox{mb}}

\def\GeV{\,\mbox{GeV}}

\def\eps{\varepsilon}

\def\Jpsi{J\!/\!\psi}
\def\psip{\psi^{\,\prime}}

\def\Y{\Upsilon}
\def\Yp{\Upsilon^{\,\prime}}
\def\Ypp{\Upsilon^{\,\prime\prime}}
\def\sqq{\sigma_{\bar QQ}}

\def\aem{\alpha_{em}}

\def\lsim{\mathrel{\rlap{\lower4pt\hbox{\hskip1pt$\sim$}}
     \raise1pt\hbox{$<$}}}         
 \def\gsim{\mathrel{\rlap{\lower4pt\hbox{\hskip1pt$\sim$}}
     \raise1pt\hbox{$>$}}}         

\begin{document}

\title{
Searching for gluon saturation effects in the
momentum transfer dependence of coherent
charmonium electroproduction off nuclei
}

\author{J. Nemchik$^{1,2}$}
\email{jan.nemcik@fjfi.cvut.cz}

\author{J. \'Obertov\'a$^{1}$}
\email{jaroslava.obertova@fjfi.cvut.cz}

\affiliation{
\vspace*{0.20cm}
$^1$
Czech Technical University in Prague, FNSPE, B\v rehov\'a 7, 11519
Prague, Czech Republic}
\affiliation{$^2$
Institute of Experimental Physics SAS, Watsonova 47, 04001 Ko\v sice, Slovakia
}

\date{\today}
\begin{abstract}

\vspace*{5mm}

We study for the first time the transverse momentum transfer 
distributions $d\sigma/dt$ in coherent production of charmonia 
in nuclear ultra-peripheral and electron-ion collisions within 
the QCD color dipole approach based on a rigorous Green function formalism. 
This allows us to treat properly the color transparency effects,
as well as the higher and leading-twist shadowing corrections
associated with the $|Q\bar Q\ra$ and $|Q\bar QnG\ra$ Fock 
components of the photon.
While the multi-gluon photon fluctuations represent the dominant 
source of nuclear shadowing at kinematic regions related to
the recent LHC and its future upgrade to LHeC, 
the upcoming electron-ion collider at RHIC will additionally require
the proper incorporation of reduced quark shadowing.
The latter effect leads to a significant decrease in the differential 
cross sections $d\sigma/dt$ compared to standard calculations 
based on the eikonal form for the dipole-nucleus amplitude.
The leading-twist shadowing corrections, corresponding to a non-linear 
QCD evolution of a partial dipole-nucleus amplitude, reduce substantially
charmonium $t$-distributions in the LHeC energy range.
We predict a non-monotonic energy dependence of $d\sigma/dt$ 
suggesting so possible gluon saturation effects with increased 
onset at larger $t$-values.
In addition to shadowing corrections, we study how the color 
transparency effects affect the shape of $t$-dependent nuclear 
modification factor.
We also briefly discuss several aspects that can modify
the charmonium production rate and thus may have a large impact 
on the search for gluon saturation effects.

\end{abstract}


\maketitle

%
%
%
\section{Introduction}
\label{intro}
%
%
%

Recent experiments with ultra-peripheral collisions (UPC) at 
Relativistic Heavy Ion Collider (RHIC) and the Large Hadron 
Collider (LHC) (see e.g., 
\cite{Bertulani:2005ru})
offer new possibilities for investigation of exclusive 
photoproduction of vector mesons on protons and nuclei. 
\cite{Bauer:1977iq}. 
%
Whereas the LHC kinematic region allows to study the space-time pattern 
of the diffraction mechanism 
\cite{Hufner:1996dr} 
in coherent photoproduction of heavy quarkonia 
at high photon energies,
the experiments on the planned Large Hadron-Electron 
Collider (LHeC) from the High Luminosity-LHC 
\cite{LHeC:2020van} 
will extend such studies also to an electroproduction process 
with a non-zero photon virtuality $Q^2$.
Here the theoretical description is simplified due to 
the long coherence length (CL) $l_c$ for the $Q\bar Q$ photon 
fluctuation, much longer than the nuclear radius $R_A$,
%
\BE 
l_c = \frac{W^2 + Q^2 -m_N^2}{m_N(M_V^2 + Q^2)} \gg R_A
\,,
  \label{lc-qq}
\EE
%
where $W$ is the c.m. energy of the photon-nucleon system; $m_N$ 
and $M_V$ are the masses of the nucleon and vector meson, respectively. 
The inequality (\ref{lc-qq}) is related to the strongest 
onset of the initial state quark shadowing.
Then the corresponding expressions for the nuclear cross sections 
lie on the eikonal form for the $Q\bar Q$ dipole-nucleus cross 
section and/or for the impact-parameter-dependent partial amplitude 
\cite{Kopeliovich:2020has,Kopeliovich:2022jwe,Nemchik:2022kkx}.
The eikonal form is frequently adopted in the literature also in kinematic regions where
the condition (\ref{lc-qq}) is not completely fulfilled (see e.g., Ref.~
\cite{Krelina:2019gee}). 

Future experiments at the Electron-Ion Collider (EIC) 
\cite{Accardi:2012qut,Aschenauer:2014cki,Aschenauer:2017jsk} 
using the present RHIC facility will provide an opportunity 
to look inside a diffraction mechanism also in the kinematic region 
where $l_c\lsim R_A$. 
Here, one cannot use the eikonal approximation anymore and more
sophisticated formalism is required to describe quantum coherence (QC) effects.  
In heavy quarkonium electroproduction, a short lifetime of the $Q\bar Q$
photon fluctuation, $l_c\ll R_A$, is correlated with a much longer 
formation time (length) (FL) 
\cite{Kopeliovich:2001xj},
$l_f\sim R_A$, controlling the evolution of the $Q\bar Q$ wave 
packet during propagation through the medium. 
The magnitude of $l_f$ can be obtained in the rest frame 
of the nucleus from the uncertainty principle, 
%
\BE
l_f = \frac{W^2+Q^2-m_N^2}{m_N (M_{V'}^2-M_V^2)}
\approx
l_c\,\cdot\frac{M_V^2+Q^2}{M_{V'}^2-M_V^2}
=
l_c\,\cdot\biggl [\frac{M_{V'}^2+Q^2}{M_{V'}^2-M_V^2} - 1\biggr]
\,,
  \label{lf-qq}
\EE
%
where $M_{V'}$ is the mass of radially excited quarkonium. 
The propagation of the $Q\bar Q$ pair in the medium is related 
to the color transparency (CT) effect (see e.g., Refs. 
\cite{Kopeliovich:1981pz,Kopeliovich:1991pu,Nemchik:2002ug,Kopeliovich:2007mm}),
which represents the final state absorption of the produced quarkonia. 
Here the medium becomes more transparent for $Q\bar Q$ dipole 
configurations with smaller transverse sizes.
Note that in the electroproduction of heavy quarkonia there is a strong inequality,
$l_c\ll l_f$, as follows from Eq.~(\ref{lf-qq}).

The description of the diffraction mechanism in terms of color 
dipoles has a long-standing history starting from Ref.~
\cite{Kopeliovich:1981pz} 
and has been applied to exclusive photo- and electroproduction of heavy 
quarkonia on protons and nuclei (see e.g., Refs.~
\cite{Kopeliovich:1991pu,Hufner:2000jb,Ivanov:2002kc,Krelina:2018hmt,Kopeliovich:2021dgx}). 
%
In our recent work 
\cite{Nemchik:2024lny}
we have studied the coherent process $\gamma^* A\to V A$ 
(V = $J/\Psi(1S)$, $\psip(2S)$) including properly the above 
mentioned effects of QC and CT within a rigorous quantum 
mechanical approach based on the Green function formalism. 
We have performed corresponding predictions for $t$-integrated\footnote{$t=-q^2$, 
where $\vec q$ is the transverse component of the momentum transfer.}
production cross sections that can be verified by future experiments at the EIC.

In Ref.~
\cite{Kopeliovich:2022jwe}, 
the $t$-dependent differential cross sections $d\sigma^{\gamma^*A\to V A}/dt$ 
have been studied in the LHC kinematic region of recent experiments with UPC, 
where the eikonal approximation related to the condition 
~(\ref{lc-qq}) can be safely adopted. 
In the present work, we improve such predictions extending them also
to the kinematic region beyond the validity of Eq.~(\ref{lc-qq}).
This requires to incorporate the reduced effects of QC by applying the path integral technique.
Simultaneously we aim to minimize all known theoretical uncertainties 
of the quantum chromodynamics (QCD) dipole formalism,
related to quarkonium S-wave functions together with elimination of D-wave component, 
dipole orientation, as well as the magnitude of the leading-twist shadowing corrections.
Consequently, this may lead to increased efficiency in searching 
for a conclusive signal of gluon saturation effects
\cite{Kovchegov:2023bvy}.

We include the correlation between the dipole transverse orientation 
$\vec r$ and the impact parameter of the collision $\vec b$ 
\cite{Kopeliovich:2007fv}.
%
This effect causes that 
the $Q\bar Q$ dipole-nucleon interaction vanishes when $\vec{r}\perp\vec{b}$ 
but reaches maximal strength when $\vec{r}\parallel\vec{b}$. However, it is missed in many calculations.
The $\vec r$-$\vec b$ correlation has been treated in Ref.~
\cite{Cepila:2025rkn}, 
but not properly incorporated, since the $\vec b$-dependent 
dipole-nucleon amplitude $\mathcal{A}_{\bar QQ}^{N}(\vec r, \vec{b}\,)$
after integration over $\vec b$ exhibits a non-monotonic behavior 
as a function of the dipole size $r$.
Besides, the ratio of dipole-nucleon amplitudes,
$R_{\perp/\parallel}(b) = 
A_{Q\bar Q}^N(\vec r,\vec b,\Theta=\pi/2) / 
A_{Q\bar Q}^N(\vec r,\vec b,\Theta=0)$ 
leads to incorrect limiting values that differ from expected
magnitudes 1 and 0
when $b\to 0$ and $b\sim R_A$
(see a brief discussion in point iii) of Sec.~\ref{pitfalls} ).

The impact of the $\vec r$-$\vec b$ correlation on the magnitude of differential 
cross sections in exclusive photo-production of heavy quarkonia on protons 
has been studied in Ref.~
\cite{Kopeliovich:2021dgx}.
%
The nodal structure of the wave functions for radially excited 
heavy quarkonium states enhances the onset of the correlation 
effect and thus provides additional constraints on the models for 
the $b$-dependent dipole amplitude. 
The importance of the color dipole orientation in other processes 
has been discussed in Ref.~
\cite{Kopeliovich:2021dgx} 
(see also references therein).
The $\vec r$-$\vec b$ correlation has been also omitted in Ref.~
\cite{Mantysaari:2024zxq}
analyzing the impact-parameter dependence in the initial condition of
the Balitsky–Kovchegov (BK) equation \cite{balitsky,kovchegov}.

However, in comparison with nucleon target,
the effect of $\vec r$-$\vec b$ correlation 
is diluted in processes on nuclear targets 
\cite{Kopeliovich:2022jwe}.
%
A rather weak impact of the dipole orientation on the azimuthal 
asymmetry of photons and pions has been found and analyzed in Refs.~
\cite{Kopeliovich:2008dy,Kopeliovich:2007sd,Kopeliovich:2007fv,Kopeliovich:2008nx}.  
%
Nevertheless, in the present paper, we implement this effect in 
calculations of $d\sigma^{\gamma^*A\to V A}/dt$, thus minimizing the 
theoretical uncertainties.

The LHC and LHeC kinematic region, related to the condition (\ref{lc-qq}),
gives rise to a maximal strength of the higher-twist quark shadowing.
However, such a shadowing correction is rather small for the photo- 
and electroproduction of heavy quarkonia due to the large heavy quark mass.
Then the main nuclear effect comes from the leading-twist gluon shadowing, 
which is related to the higher Fock components of the projectile photon 
containing gluons.
According to the analysis in Ref.~
\cite{Kopeliovich:2022jwe},
the dominant contribution to nuclear shadowing comes from the $|Q\bar QG\ra$ 
Fock component.
We implement the gluon shadowing (GS) corrections 
in our calculations of nuclear cross sections from Refs.~
\cite{Ivanov:2002kc,Kopeliovich:1999am,Kopeliovich:2022jwe}.
%
Inclusion of higher multi-gluon Fock components $|Q\bar QnG\ra$ is still a challenge. 
However, their effect is essentially taken into account by the 
eikonalization of the GS correction factor.

The LHeC kinematic region allows to analyze how a modification 
of the gluon distribution in nuclei by the GS corrections 
affects the $t$-dependent differential cross sections.
In the infinite momentum frame, the phenomenon of GS looks similar 
to gluon-gluon fusion corresponding to a non-linear term in evolution equations
\cite{Gribov:1981ac,Gribov:1983ivg,balitsky,kovchegov}.
We expect a suppression of small-$x$ gluons 
and a precocious onset of saturation effects, especially for heavy nuclei. 
This may lead to a non-monotonic energy dependence 
of $d\sigma^{\gamma^*A\to V A}/dt$ at fixed $t$-values.
Such an expectation is explored in the present work and represents
one of the main goals of our study.
Note that a non-monotonic energy dependence of $d\sigma/dt$ has been presented 
at large $t\sim 1\,\GeV^2$ also in Ref.~
\cite{Cepila:2023dxn} 
studying the incoherent production of $\Jpsi$ on the lead 
target within the hot-spot model.
However, the important effect of gluon shadowing has been ignored. 
This may have a large impact on the reliability of the predicted 
onset of saturation effects.

The present paper is organized as follows. 
In the next Sec.~\ref{dipole-p} we briefly present the dipole formalism 
for electroproduction on a proton target, $\gamma^*p\to V p$, 
in terms of the $\vec b$-dependent dipole amplitude. 
The Sec.~\ref{dipole-A} is devoted to the coherent heavy quarkonium 
production on nuclear targets, $\gamma^*A\to V A$.
Here we present expressions for the $t$-dependent differential cross sections
$d\sigma^{\gamma^*A\to V A}/dt$ within a rigorous Green function formalism.
Consequently, in Sec.~\ref{data} we first compare our model calculations 
of $d\sigma^{\gamma^*p\to V p}/dt$ ($V=\Jpsi(1S)$ and $\psip(2S)$)
and $\sigma^{\gamma p\to \psip(2S) p}$ with available data from 
experiments at the Hadron–Electron Ring Accelerator (HERA) 
and the LHC.
Then in Sec.~\ref{CT} we propose how manifestations of CT effects may be 
recognized by the future EIC measurements. 
Sec.~\ref{quark-shadowing} is devoted to predictions for $d\sigma/dt$ 
in the coherent photo- and electroproduction of 1S and 2S charmonium states off nuclei
in kinematic regions accessible by upcoming EIC experiments at RHIC and the LHeC.
Here we analyze the impact of reduced quantum coherence effects 
for $Q\bar Q$ photon fluctuations on magnitudes of the $t$-dependent 
nuclear differential cross sections at small photon energies $W$, 
when $l_c\lsim R_A$, for fixed values of $Q^2$ and $t$.
Moreover, for the LHeC kinematic region, we predict in 
Sec.~\ref{gluon-shadowing} the possible onset of saturation effects, 
manifested via a non-monotonic energy behavior of $d\sigma^{\gamma^*A\to V A}/dt$.
Finally, in Sec.~\ref{pitfalls} we briefly mention and discuss 
possible pitfalls in searching for gluon saturation effects.
The last Sec.~\ref{sec-sum} contains summary and discussion of our results.

%
%
%
\section{Electro-production of heavy quarkonia on protons}
\label{dipole-p}
%
%
%

Within the light-front (LF) color dipole formalism, the amplitude 
for the electroproduction of heavy quarkonia with the transverse 
component of momentum transfer $\vec q$ 
takes the following factorized form
\cite{Kopeliovich:1991pu},
%
\BA
\mathcal{A}^{\gamma^\ast p\to V p}(x,Q^2,\vec q)
=
\bigl\la V |\tilde{\mathcal{A}} |\gamma^*\bigr\ra
=
\int d^2r\int_0^1 d\alpha\,
\Psi_{V}^{*}(\vec r,\alpha)\,
\mathcal{A}_{Q\bar Q}(\vec r, x, \alpha,\vec q)\,
\Psi_{\gamma^\ast}(\vec r,\alpha,Q^2)\,.
  \label{amp-p0}
\EA
%
Here
$\mathcal{A}_{Q\bar Q}(\vec r, x, \alpha,\vec q)$ is the amplitude for 
the elastic scattering of the color dipole on a nucleon target.

It is convenient to study differential cross sections $d\sigma/dq^2$ 
treating the partial dipole scattering amplitude  
in the impact parameter representation $\mathcal{A}_{Q\bar Q}(\vec r, x, \alpha,\vec b)$
related to the $\vec q$-dependent amplitude by Fourier transform,
%
\BA
\mathcal{A}_{Q\bar Q}(\vec r, x, \alpha,\vec q)
=
\int d^2 b\,e^{\, i\, \vec b\cdot\vec q}\,
\mathcal{A}_{Q\bar Q}(\vec r, x, \alpha,\vec b)\,,
\EA
%
which correctly reproduces the dipole cross section at $\vec q=0$,
%
\BA
\sqq(r,x)
=
\mathrm{Im}\mathcal{A}_{Q\bar Q}(\vec r, x, \alpha,\vec q=0) 
= 
2\,\int d^2 b\,
\mathrm{Im}\mathcal{A}^N_{Q\bar Q}(\vec r, x, \alpha,\vec b)\,.
  \label{dcs-b}
\EA
%

In Eq.~(\ref{amp-p0}), the variable
$\Psi_V(r,\alpha)$ is the LF wave function for heavy quarkonium and
$\Psi_{\gamma^\ast}(r,\alpha,Q^2)$ is the LF distribution
of the $Q\bar Q$ Fock component of the real ($Q^2 = 0$) or virtual ($Q^2 > 0$) photon,
with transverse separation $\vec{r}$.
The variable $\alpha$ is the fractional LF momentum carried by a heavy quark or antiquark from a $Q\bar Q$ Fock component of the photon.

Combining Eqs.~(\ref{amp-p0})-(\ref{dcs-b}) the electroproduction amplitude    
$\mathcal{A}^{\gamma^{\ast} p\to V p}(x,Q^2, \vec q)$ has the following
form,
%
\BA
\!\!\!\!\!\!
\mathcal{A}^{\gamma^{\ast} p\to V p}(x,Q^2, \vec q)
=
2\,
\int d^2 b\,
e^{\, i\, \vec b\cdot\vec q}\,
\int d^2r\int\limits_0^1 d\alpha\,
\Psi_{V}^{*}(\vec r,\alpha)\,
\mathrm{Im} 
\mathcal{A}^N_{Q\bar Q}(\vec r, x, \alpha,\vec b)\,
\Psi_{\gamma^{\ast}}(\vec r,\alpha,Q^2)\,,
  \label{amp-p}
\EA
%
%
where the impact parameter $\vec b$ of
the dipole is the transverse distance from the target to
the dipole center of gravity, which 
varies with the fractional LF momenta
of $Q$ or $\bar Q$.

The amplitude (\ref{amp-p}) depends on Bjorken $x$ evaluated in 
\cite{ryskin} 
in the leading $\log(1/x)$ approximation, 
\BE
x=\frac{M_V^2+Q^2}{s}=
\frac{M_{V}^2+Q^2}{W^2+Q^2-m_N^2}\,.
  \label{x}
\EE

The essential feature of the  dipole-proton partial amplitude  
$\mathcal{A}^N_{\bar QQ}(\vec r, x, \alpha,\vec b\,)$ in Eq.~(\ref{amp-p}) 
is the $\vec r$-$\vec b$ correlation.
Its explicit form was proposed in Refs.~
\cite{Kopeliovich:2008dy,Kopeliovich:2007sd,Kopeliovich:2007fv,Kopeliovich:2008nx,Kopeliovich:2021dgx} 
and is as follows,
%
\BA
\mathrm{Im} \mathcal{A}^N_{\bar QQ}(\vec r, x, \alpha,\vec b\,)
=
\frac{\sigma_0}{8\pi \mathcal{B}(x)}\,
\Biggl\{
\exp\left[-\,\frac{\bigl [\vec b+\vec
r(1-\alpha)\bigr ]^2}{2\mathcal{B}(x)}\right] 
+ 
\exp\left[-\,\frac{(\vec
b-\vec r\alpha)^2}{2\mathcal{B}(x)}\right]
\nonumber\\
- \,2\,\exp\Biggl[-\,\frac{r^2}{R_0^2(x)}
-\,\frac{\bigl [\,\vec b+(1/2-\alpha)\vec
r\,\bigr ]^2}{2\mathcal{B}(x)}\Biggr]
\Biggr\}\,,
  \label{dipa-gbw}
 \EA
%
where the function $\mathcal{B}(x)$ was defined in Ref.~
\cite{Kopeliovich:2008nx}
as,
\BE
 \mathcal{B}(x)=B^{\bar qq}_{el}(x, r\to 0)-
 \frac{1}{8} R_0^2(x)\,.
  \label{slope}
 \EE
Here $R_0^2(x)$ controls the $x$ dependence of the saturation cross section, introduced in 
\cite{GolecBiernat:1998js,GolecBiernat:1999qd},
$\sqq(r,x)=\sigma_0\,\bigl (1 - \exp \bigl [ - {r^2}/{R_0^2(x)}\bigr ] \bigr )$.
We adopt parameters from the recent work
\cite{Golec-Biernat:2017lfv},
$\sigma_0 = 25.21\,\mb$, $R_0(x) = 0.4\,\fm\times(x/x_0)^{0.1405}$ with $x_0 = 0.80\times 10^{-4}$,
referred to as the GBS-0 model.

The dipole-proton slope in the limit of vanishingly 
small dipoles $B^{\bar qq}_{el}(x, r\to 0)$ can be 
measured in the electroproduction of vector mesons 
with highly virtual photons $Q^2\gg 1\GeV^2$. 
The measured slope $B_{\gamma^*p\to\rho p}(x,Q^2\gg 1\GeV^2)\approx 5\GeV^{-2}$ 
\cite{ZEUS:2007iet} 
is, as expected
\cite{jan-98},
defined by the proton charge radius.

The GBS-0 model mentioned above lacks the Dokshitzer-Gribov-Lipatov-Altarelli-Parisi
(DGLAP) evolution.
However, Ref.~
\cite{Golec-Biernat:2017lfv}
also contains 
the same model with such an evolution.
Here the saturation scale is related to the gluon density, 
which is subject to the DGLAP evolution, 
$R_0^2(x,\mu^2) = {4}/{Q_s^2(x,\mu^2)}={\sigma_0\,N_c} / 
\bigl ({\pi^2\,\alpha_s(\mu^2)\,x\,g(x,\mu^2)}\bigr )$, 
where $\mu^2 = {\mathrel{C}} / {r^2} + \mu_0^2$, $Q_s^2$ 
is the saturation scale,
and the gluon distribution function $x\,g(x,\mu^2)$ is obtained 
as a solution of the DGLAP evolution equation with the form 
at the initial scale $Q_0^2 = 1\,\GeV^2$, $x\,g(x,Q_0^2)=A_g \,x^{-\lambda_g} (1-x)^{5.6}$. 
Here $A_g = 1.07$, $\lambda_g = 0.11$, $\mu_0^2 = 1.74\,\GeV^2$, 
$\mathrel{C} = 0.27$ and $\sigma_0 = 22.93\,\mb$. 
In what follows, we will refer to the GBS dipole model as to the GBS-0 model 
containing the DGLAP evolution and the color dipole orientation (\ref{dipa-gbw}).

Treating the electroproduction of heavy quarkonia and 
assuming the $s$-channel helicity conservation,
the $t$-dependent differential cross section 
can be expressed as a sum of $T$ and $L$ contributions 
for transversely and longitudinally polarized photons and vector mesons, respectively,
%
\BA
\frac{d\sigma^{\gamma^{\ast}p\rightarrow Vp}(x,Q^2,t=-q^2)}{dt}
&=&
\frac{d\sigma_T^{\gamma^{\ast}p\rightarrow Vp}(x,Q^2,t)}{dt} +
\tilde{\eps}\,
\frac{d\sigma_L^{\gamma^{\ast}p\rightarrow Vp}(x,Q^2,t)}{dt}
\nonumber\\
&=&
\frac{1}{16\pi}\left(\Big\vert \mathcal{A}^{\gamma^{\ast}p\rightarrow Vp}_{T}(x,Q^2,\vec q~)\Big\vert^{2}+
\tilde{\eps}\Big\vert \mathcal{A}^{\gamma^{\ast}p\rightarrow Vp}_{L}(x,Q^2,\vec q~)\Big\vert^{2}\right) \,,
  \label{total-cs}
\EA
%
where we have taken the photon polarization $\tilde{\eps} = 0.99$.

We also include a small real part 
\cite{Bronzan:1974jh,Nemchik:1996cw,Forshaw:2003ki} 
of the $\gamma^{\ast} p\to V p$ amplitude
performing the following replacement in Eq.~(\ref{amp-p}),
%
\BA
\mathrm{Im} \mathcal{A}_{\bar QQ}^{N}(\vec r, x, \alpha, \vec{b}\,)
\Rightarrow
\mathrm{Im} \mathcal{A}_{\bar QQ}^{N}(\vec r, x, \alpha, \vec{b}\,)
\,\cdot
\left(1 - i\,\frac{\pi\,\Lambda}{2}\right)\,,
\qquad\qquad
\Lambda
=
\frac
{\partial
 \,\ln\,({\mathrm{Im}\mathcal{A}_{\bar QQ}^{N}(\vec r, x, \alpha, \vec{b}\,))}}
{\partial\,\ln (1/x)}\ .
  \label{re/im}
\EA
%

In order to include the skewness correction 
\cite{Shuvaev:1999ce} 
we perform the following modification,
%
\BA
\mathrm{Im} \mathcal{A}^N_{Q\bar Q}(\vec r, x, \alpha,\vec{b})\Rightarrow
\mathrm{Im} \mathcal{A}^N_{Q\bar Q}(\vec r, x, \alpha,\vec{b})\cdot 
R_S(\Lambda)\,
  \label{skewness-a}
\EA
%
where the skewness factor 
$R_S(\Lambda)=\bigl ({2^{2\,\Lambda + 3}}/{\sqrt{\pi}}\, \bigr) \,\cdot {\Gamma(\Lambda + 5/2)}/{\Gamma(\Lambda + 4)}$.

In our previous works 
\cite{Krelina:2018hmt,Cepila:2019skb,Krelina:2019egg,Kopeliovich:2020has,Krelina:2020bxt,Kopeliovich:2021dgx,Nemchik:2024lny} 
(see also Sec. II.A in Ref.~\cite{Kopeliovich:2022jwe})
we ruled out the possibility of a photon-like vertex for the transition 
of a heavy quankonium to a $Q\bar Q$ pair due to the unusually large weight 
of a $D$-wave component in the rest frame wave function, 
in contrast to solutions of the Schr\"odinger equation.
Instead, we rely on the quarkonium LF wave functions obtained from a solution of the Lorentz boosted Schr\"odinger equation 
\cite{Kopeliovich:2015qna} 
with realistic potentials. 
Although several such potentials can be found in the literature, in the present paper we choose
only the power-like potential (POW) 
\cite{Barik:1980ai}.
This potential 
in combination with the GBS model 
\cite{Golec-Biernat:2017lfv} 
for the dipole cross section 
provides the best description of the UPC data, as shown in Ref.~
\cite{Nemchik:2024lny}.
%
Since the quark transverse momenta are not parallel to the boost axis, a significant 
correction to the boosting prescription 
\cite{Terentev:1976jk} 
from the rest frame to the LF frame is taken into account,
known as the Melosh spin rotation 
\cite{Melosh:1974cu} 
(see also 
\cite{Krelina:2018hmt,Lappi:2020ufv}).
%
This leads  
to the following specific form of the electroproduction amplitudes 
given by Eq.~(\ref{amp-p}) for $T$ and $L$ polarizations,
%
\BA
\mathcal{A}_T^{\gamma^{\ast} p\to V p}(x,Q^2,\vec{q})
&=&
N_p\,
\int d^2r
\int_0^1 d\alpha  \,
\int d^2 b\,
e^{\, i\, \vec b\cdot\vec q}\,\,
\mathrm{Im} 
\mathcal{A}^N_{Q\bar Q}(\vec r, x, \alpha,\vec b)\,
\left[\Sigma^{(1)}_{T}(r,\alpha,Q^2) 
+
\Sigma^{(2)}_{T}(r,\alpha,Q^2)\right]\,,
 \nonumber
\\ 
&\mbox{with}&
\nonumber\\
\!\!\!\!\!\!\!\!\!\!
   \Sigma^{(1)}_T(r,\alpha,Q^2)
   &=&  
   K_0(\eta r) \int_0^\infty dp_T\,p_T\,
   J_0(p_T r) \Psi_V (\alpha,p_T) 
   \left[ \frac{2\,m_Q^2(m_L+m_T)+m_L\,p_T^2}{ m_T (m_L + m_T)} \right]\,,
\nonumber \\
\!\!\!\!\!\!\!\!\!\!
  \Sigma^{(2)}_T(r,\alpha,Q^2)
  &=& 
   K_1(\eta r) \int_0^\infty dp_T\,p_T^2\,
   J_1(p_T r) \Psi_{V} (\alpha,p_T) \left[
\,\eta\, 
   \frac{m_Q^2(m_L+2m_T)-m_T\,m_L^2}{m_Q^2\,m_T (m_L+m_T)} \right]\,,
   \label{amp-p-sr1}
\EA
%
and 
%
\BA
\mathcal{A}_L^{\gamma^{\ast} p\to V p}(x,Q^2,\vec{q})
&=&
N_p\,
\int d^2r
\int_0^1 d\alpha  \,
\int d^2 b\,
e^{\, i\, \vec b\cdot\vec q}\,\,
\mathrm{Im} 
\mathcal{A}^N_{Q\bar Q}(\vec r, x, \alpha,\vec b)\,
\Sigma_{L}(r,\alpha,Q^2)\,,
 \nonumber
\\ 
&\mbox{with}&
\nonumber\\
\!\!\!\!\!\!\!\!\!\!
   \Sigma_L(r,\alpha,Q^2)
   &=&  
   K_0(\eta r) \int_0^\infty dp_T\,p_T\,
   J_0(p_T r) \Psi_V (\alpha,p_T) 
   \left[ 4\,Q\,\alpha\,(1-\alpha)\,\frac{m_Q^2 + m_L m_T}{m_Q\,(m_L+m_T)}
   \right]\,,
   \label{amp-p-sr2}
\EA
%
where $N_p=Z_Q\,\sqrt{2 N_c \,\alpha_{em}}/2\,\pi$;
$\eta^2 = m_Q^2\,+ \alpha (1-\alpha) Q^2$; $m_Q$ is the heavy quark mass;
$\aem= 1/137$ is the fine-structure constant; the factor $N_c=3$ represents the number of colors in QCD; $Z_Q=2/3$ is the charge isospin factor for charmonium production;  
variables
$m_T = \sqrt{m_Q^2 + p_T^2}$
and 
$m_L = 2\, m_Q\,\sqrt{\alpha(1-\alpha)}$\,,
and
$J_{0,1}$ and $K_{0,1}$ are the Bessel functions of the first kind and the modified Bessel functions of the second kind, respectively.

%
%
%
\section{
Coherent electroproduction off nuclei: 
Momentum transfer dependence in
the Green function formalism}
\label{dipole-A}
%
%
%

In this Section we focus on
the lowest $|Q\bar Q\ra$ Fock component of the projectile photon. 
Since the $Q\bar Q$ transverse size is small, $\propto 1/\eta$, 
the shadowing corrections are as small as $1/\eta^2$, 
so they should be treated as a higher-twist effect.
Within the Green function formalism, 
the amplitude for quarkonium electroproduction on a nuclear target, $\gamma^*A\to VA$, is
given by the following expression,
%
\BE
\mathcal{A}^{\gamma^{\ast} A\to V A}(x,Q^2,\vec q\,)
=
2\,
\int d^2 b\,
\int d^2b_A\,\,
\exp\Bigl [\, i\,(\,\vec{b} + \vec b_A)\,\cdot\vec{q}\,\Bigr ]
\int_{-\infty}^{\infty} dz\,
\rho_A\bigl (\vec b+\vec b_A,z\bigr )\,F_1(x,\vec b,\vec b_A,z)\,,
  \label{amp-A0}
\EE
where
$\rho_A(b,z)$ is the nuclear density {distribution}, 
for which we employ the realistic Wood-Saxon form with parameters taken from Ref.
\cite{DeJager:1987qc};
$\vec b_A$ is the nuclear impact parameter
and the function $F_{1}(x,\vec b,\vec b_A,z)$ can be expressed as,
 \BA
F_1(x,\vec b,\vec b_A,z) 
&=&
\int_0^1 d\alpha\,
\int d^{2} r_{1}\,d^{2} r_{2}\,
\Psi^{*}_{V}(\vec r_{2},\alpha)\,
G_{Q\bar Q}(z^{\prime}\to\infty,\vec r_{2};z,\vec r_{1}|\vec b_A)\,
\mathcal{A}^N_{\bar QQ}(\vec r_1, x, \alpha,\vec b)\,
\Psi_{\gamma^*}(\vec r_{1},
\alpha,Q^2)\,.
  \label{f1}
 \EA
Here the function $F_1(x,\vec b,\vec b_A,z)$ 
describes the coherent (elastic) production of 
the colorless $Q\bar Q$ pair of initial separation $\vec r_1$ at point $z$
with its subsequent evolution during propagation through the nucleus 
which is completed by the formation of
the heavy quarkonium wave function at $z^{\,\prime}\to\infty$ with final separation $\vec r_2$.
Such an evolution 
of an interacting $Q\bar Q$ pair is described by the corresponding Green function
$G_{Q\bar Q}(z^\prime,\vec r_2;z,\vec r_1|\vec b_A)$, which
satisfies
the two-dimensional Schr\"odinger equation
\cite{Kopeliovich:1999am,Kopeliovich:2001xj}
%
 \BE
i\frac{d}{dz_2}\,G_{Q\bar Q}(z_2,\vec r_2;z_1,\vec r_1|\vec b_A)=
\left[\frac{\eta^{2} - \Delta_{r_{2}}}{2\,p\,\alpha\,(1-\alpha)}
+V_{Q\bar Q}(z_2,\vec r_2,\alpha,\vec b_A)\right]
G_{Q\bar Q}(z_2,\vec r_2;z_1,\vec r_1|\vec b_A)\ ,
  \label{schroedinger}
 \EE
%
where $p$ is the photon energy in the target rest frame and the Laplacian $\Delta_{r_2}$ acts on the coordinate $r_2$.

The real part of the LF potential $V_{Q\bar Q}(z_2,\vec r_2,\alpha)$
in Eq.~(\ref{schroedinger})
describes the interaction between the $Q$ and $\bar{Q}$.
In Ref.~
\cite{Nemchik:2024lny}
we have proposed a procedure how to obtain ${\mathcal Re} V_{Q\bar Q}(z_2,\vec r_2,\alpha)$ in the LF frame from various realistic $Q\bar Q$ interaction models defined in the rest frame. 
Such potentials lead to a correct shape
of quarkonium wave functions in the LF frame. 
As in the case of the nucleon target, for the coherent 
process $\gamma^*A\to V A$ we also adopt the POW model 
\cite{Barik:1980ai}
for the realistic interaction potential in the $Q\bar Q$ rest frame.

The imaginary part of the the LF potential $V_{Q\bar Q}(z_2,\vec r_2,\alpha)$ 
in Eq.~(\ref{schroedinger}) controls the attenuation of the $|Q\bar Q\ra$ Fock state
of the photon in the medium. 
It has the following form,
%
 \BE
{\mathcal Im} V_{Q\bar Q}(z_2,\vec r,\alpha,\vec b_A) 
= 
-
\int d^2 b'
\mathcal{A}^N_{\bar QQ}(\vec r, x, \alpha,\vec b\,')\,
\rho_{A}(\vec b\,'+\vec b_A,z_2)\,.
  \label{pot-im}
 \EE 
%

The final form of the nuclear $t$-dependent production amplitudes for both the T and L polarizations, including the spin rotation effects, 
has a structure that is more complicated compared to the proton target, 
Eqs.~ (\ref{amp-p-sr1}) and (\ref{amp-p-sr2}), and reads,
%
%
\BA
\!\!\!\!\!\!\!\!\!\!
&&
\mathcal{A}_T^{\gamma^{\ast} A\to V A}(x,Q^2,\vec{q})
=
2\,N_p\,
\int d^2 b\,
\int d^2b_A\,\,
\exp\Bigl [\, i\,(\,\vec{b} + \vec b_A)\,\cdot\vec{q}\,\Bigr ]
\int_{-\infty}^{\infty} dz\,
\rho_A\bigl (\vec b+\vec b_A,z\bigr )\,
\nonumber\\
&\times&
\int d^2 r_1\,
\int d^2 r_2\,
\int_0^1 d\alpha\,
\mathcal{A}^N_{Q\bar Q}(\vec r_1, x, \alpha,\vec b)\,
G_{Q\bar Q}(z^{\prime}\to\infty,\vec r_{2};z,\vec r_{1}|\vec b_A)\,
\left[\Gamma^{(1)}_{T}(r_1,r_2,\alpha,Q^2) 
+
\Gamma^{(2)}_{T}(r_1,r_2,\alpha,Q^2)\right]\,,
 \nonumber
 \\
\mbox{with}
\nonumber\\
&&   \Gamma^{(1)}_T(r_1,r_2,\alpha,Q^2)
=
   K_0(\eta r_1) 
   \int_0^\infty dp_T\,p_T\,
   J_0(p_T r_2) \Psi_V (\alpha,p_T) 
   \left[ \frac{2\,m_Q^2(m_L+m_T)+m_L\,p_T^2}{ m_T (m_L + m_T)} \right]\,,
\nonumber \\
&&  \Gamma^{(2)}_T(r_1,r_2,\alpha,Q^2)
=
   K_1(\eta r_1) 
   \int_0^\infty dp_T\,p_T^2\,
   J_1(p_T r_2) \Psi_{V} (\alpha,p_T) 
   \left[\,\eta\, \frac{m_Q^2(m_L+2m_T)-m_T\,m_L^2}{m_Q^2\,m_T (m_L+m_T)} \right]\,,
   \label{amp-A-sr1}
\EA
%
and 
%
\BA 
&&
\mathcal{A}_L^{\gamma^{\ast} A\to V A}(x,Q^2,\vec{q})
=
2\,N_p\,
\int d^2 b\,
\int d^2b_A\,\,
\exp\Bigl [\, i\,(\,\vec{b} + \vec b_A)\,\cdot\vec{q}\,\Bigr ]
\int_{-\infty}^{\infty} dz\,
\rho_A\bigl (\vec b+\vec b_A,z\bigr )\,
\nonumber\\
&\times&
\int d^2 r_1\,
\int d^2 r_2\,
\int_0^1 d\alpha\,
\mathcal{A}^N_{Q\bar Q}(\vec r_1, x, \alpha,\vec b)\,
G_{Q\bar Q}(z^{\prime}\to\infty,\vec r_{2};z,\vec r_{1}|\vec b_A)\,
\Gamma_{L}(r_1,r_2,\alpha,Q^2) \,,
\nonumber
\\ 
\mbox{with}
\nonumber\\
&&
   \Gamma_L(r_1,r_2,\alpha,Q^2)
=
   K_0(\eta r_1) \int_0^\infty dp_T\,p_T\,
   J_0(p_T r_2) \Psi_V (\alpha,p_T) 
   \left[ 4\,Q\,\alpha\,(1-\alpha)\,\frac{m_Q^2 + m_L m_T}{m_Q\,(m_L+m_T)}
   \right]\,.
   \label{amp-A-sr2}
\EA
%
The expression for the $t$-dependent differential cross section reads
%
\BE
\frac{d\sigma^{\gamma^\ast A\to V A}(x,Q^2,t=-q^2\,)}{dt}
=
\frac{1}{16\pi}
\left(\Big\vert \mathcal{A}^{\gamma^{\ast}A\rightarrow VA}_{T}(x,Q^2,\vec q~)\Big\vert^{2}+
\tilde{\eps}\,\Big\vert \mathcal{A}^{\gamma^{\ast}A\rightarrow VA}_{L}(x,Q^2,\vec q~)\Big\vert^{2}\right) \,.
\label{nucleus}
\EE

Note that the Green function formalism can be applied to coherent processes without any restrictions for the coherence length $l_c$, for arbitrary realistic $Q$-$\bar Q$ interaction potential, as well as for any phenomenological model for the impact parameter-dependent partial dipole amplitude.

At high photon energies, corresponding to $x\lsim 0.01$,
the condition (\ref{lc-qq}) is valid and the so-called
{\sl long coherence length} (LCL) regime is at work. Then the 
eikonal approximation can be applied as a limiting case of the Green function formalism, 
when the Green function acquires a simple form, 
%
 \BE
G_{Q\bar Q}(z_2,\vec r_2;z_1,\vec r_1) \Rightarrow
\delta^{(2)}(\vec r_1-\vec r_2)\,\exp\left[
- \,
\int d^2 b\,
\mathcal{A}^N_{Q\bar Q}(\vec r_1, x, \alpha,\vec b\,)\,
\int\limits_{z_1}^{z_2} dz\,\rho_A(\vec b+\vec b_A,z)\right]\, ,
  \label{gf-lcl}
 \EE
%
and the Lorentz time dilation freezes the transverse sizes 
of such long-lived  $Q\bar Q$ photon fluctuations 
during propagation through the medium.
Such an LCL effect is also known as the "frozen" approximation. 
Then
the $t$-dependent amplitude of quarkonium electroproduction on a nuclear target, 
$\gamma^*A\to VA$ is given by the expression (\ref{amp-p}), but replacing 
the dipole-nucleon by dipole-nucleus amplitude 
$\mathcal{A}^N_{\bar QQ}(\vec r, x, \alpha,\vec b\,)\Rightarrow 
\mathcal{A}^A_{\bar QQ}(\vec r, x, \alpha,\vec b_A)$ and $\vec b\Rightarrow \vec b_A$,
\BE
\mathcal{A}^{\gamma^{\ast} A\to V A}(x,Q^2,\vec q\,)
=
2\,
\int d^2b_A\,e^{i\vec q\cdot\vec b_A}
\int d^2r\int_0^1 d\alpha\,
\Psi_{V}^{*}(\vec r,\alpha)\,
\mathcal{A}^A_{\bar QQ}(\vec r, x, \alpha,\vec b_A)\,
\Psi_{\gamma^\ast}(\vec r,\alpha,Q^2)\,,
  \label{amp-A0m}
\EE
where one can rely on
the eikonal form for the dipole-nucleus partial amplitude at nuclear impact parameter $\vec {b}_A$, 
\BA
\mathrm{Im} \mathcal{A}^A_{\bar QQ}(\vec r, x, \alpha,\vec b_A)\Biggl |_{l_c\gg R_A}
&=&
1 - \Biggl [1 - \frac{1}{A}\,
\int d^{2} b\,\,
\mathrm{Im} \mathcal{A}^N_{\bar QQ}(\vec r, x, \alpha, \vec{b}\,)\,
T_{A}(\vec{b}_A+\vec{b}\,)
\Biggr ]^A
\nonumber\\
&\approx&
1 - \exp \Biggl [ - \int d^{2} b\,\,
\mathrm{Im} \mathcal{A}^N_{\bar QQ}(\vec r, x, \alpha, \vec{b}\,)\,
T_{A}(\vec{b}_A+\vec{b}\,)
\Biggr ]
\,.
  \label{eik}
\EA
Here $T_A(\vec b_A) = \int_{-\infty}^{\infty} dz\,\rho_A(\vec b_A,z)$ is the nuclear thickness function normalized as $\int d^2 b_A\,T_A(\vec b_A) = A$.


Besides the effect of the reduced coherence length when $l_c\lsim (1\div 2)\cdot R_A$, 
the photoproduction of heavy quarkonia is also affected  by the gluon shadowing 
as shown in
Refs.~\cite{Ivanov:2007ms, Kopeliovich:2020has,Kopeliovich:2022jwe,Nemchik:2024lny}.
%
The leading-twist gluon shadowing was introduced in 
Ref.~\cite{Kopeliovich:1999am} 
within the dipole representation
as a shadowing correction corresponding to higher Fock
components of the photon containing gluons, 
i.e., $|Q\bar QG\ra$, $|Q\bar Q2G\ra$, ... ,$|Q\bar QnG\ra$.
However, in 
agreement
with the analysis and discussion in 
Ref.~\cite{Kopeliovich:2022jwe},
the high-energy part of the EIC kinematic region at RHIC 
and UPC kinematic region at the LHC
generate
a dominant contribution to the nuclear shadowing coming only from
one-gluon Fock state $|Q\bar QG\ra$ of the photon.
The corresponding coherence (radiation) length has the form
\BE
l_c^G
=
\frac{(W^2 + Q^2 - m_N^2)\cdot\alpha_g(1-\alpha_g)}
{m_N\Bigl [k_T^2+(1-\alpha_g)\,m_g^2+\alpha_g M_{\bar QQ}^2 + \alpha_g (1-\alpha_g)\,Q^2\Bigr] }, 
  \label{lg-full}
\EE
where $\alpha_g$ is the LF fraction of the photon momentum carried by the gluon,
$M_{\bar QQ}$ is the effective mass of the ${\bar QQ}$ pair and 
the effective gluon mass  $m_g\approx0.7\GeV$ is fixed by data on gluon radiation 
\cite{Kopeliovich:1999am,Kopeliovich:2007pq}. 
%
The condition for the onset of GS is a sufficiently long $l_c^G\gg d$, 
where $d\approx 2\,\fm$ is the mean separation between bound nucleons. 
One can also see that $l_c^G\ll l_c$ due to small $\alpha_g\ll 1$, in 
agreement with calculations in Ref.~
\cite{Kopeliovich:2000ra}.
%
We have incorporated the GS correction in our calculations as 
a reduction of the partial $b$-dependent dipole amplitude in nuclear reactions with respect 
to processes on nucleon
\cite{Kopeliovich:2001ee,Kopeliovich:2022jwe},
%
\BE
\mathrm{Im} \mathcal{A}^N_{\bar QQ}(\vec r, x, \alpha,\vec b\,)
\Rightarrow 
\mathrm{Im} \mathcal{A}^N_{\bar QQ}(\vec r, x, \alpha,\vec b\,)
\cdot R_G(x,|\,\vec b_A +\vec b\,|)\,,
  \label{eq:dipole:gs:replace-b}
\EE
where
the correction factor $R_G(x,b)$,
related to the $Q\bar QG$ component of the photon, was calculated 
within the Green function formalism 
\cite{Kopeliovich:1999am,Kopeliovich:2001xj,Kopeliovich:2001ee,Ivanov:2002kc,Nemchik:2002ug,Kopeliovich:2008ek,Krelina:2020ipn,Nemchik:2024lny}
(see also Fig.~1 in Ref.~ 
\cite{Kopeliovich:2022jwe}).
\\

\BF
\vspace*{0.5cm}
\includegraphics[height=6.50cm]{GS-bdep.eps}
\hspace*{0.70cm}
\includegraphics[height=6.50cm]{GS-int-x.eps}
\\
\Caption{
\label{Fig-gs}
    Gluon shadowing factor $R_G$ for photoproduction of $\Jpsi$ on the gold target
    as function of impact parameter $b_A$ at several fixed values of $x$ (left), and
    as function of $x$ at fixed values of the photon virtuality 
    $Q^2 = 0$ and $50\,\GeV^2$ (right). 
  }
\EF

In the left panel of Fig.~\ref{Fig-gs} we present the $b_A$ dependence of the $R_G$ factor at
several fixed values of $x$ covering the LHC and LHeC kinematical regions. 
The right panel of Fig.~\ref{Fig-gs} shows the $b_A$-integrated 
gluon shadowing factor at fixed values of the photon virtuality
$Q^2 = 0$ and $50\,\GeV^2$.

The real part of the nuclear $\gamma^{\ast} A\to \Jpsi A$ amplitude and the corresponding skewness correction has been included in consistency with Eqs.~(\ref{re/im}) and (\ref{skewness-a}). Here we checked that
final results for $t$-dependent differential cross section of the coherent process $\gamma^{\ast} Au\to\Jpsi Au$ are insensitive to the order in which such corrections are implemented, whether after the substitution (\ref{eq:dipole:gs:replace-b}) or first without it with subsequent multiplication
by the $R_G$ factor.

Note that the calculations of higher multi-gluon
Fock components are very complex within the
Green function formalism and represent thus a challenge.
However, using renormalization (\ref{eq:dipole:gs:replace-b}), 
the corresponding shadowing corrections are included essentially
via eikonalization of the factor $R_G(x,b)$ (see discussion in Ref.~
\cite{Kopeliovich:2001ee}).
\\

\BF
%
\includegraphics[height=7.0cm]{dsdt_psi1S_W-Q2_0-tfix.eps}
\hspace*{0.50cm}
\includegraphics[height=7.0cm]{dsdt_psi1S_W-Q2_8,9-tfix.eps}
\\
\vspace*{1.10cm}
\includegraphics[height=7.0cm]{sig_psi2S_W-Q2_0-tint.eps}
\hspace*{0.5cm}
\vspace*{0.0cm}
\includegraphics[height=7.0cm]{dsdt_psi1S_Pb_W125_Q2_0.eps}
\\
\vspace*{0.0cm}
\Caption{
\label{Fig1-proton}
    Model calculations of
    $d\sigma^{\gamma(\gamma^{\ast}) p\to \Jpsi p}(t)/dt$ at $Q^2=0.05\,\GeV^2$ (upper left panel) and
    at $Q^2 = 8.9\,\GeV^2$ (upper right panel) as a function of photon energy $W$
    at several fixed values of $t$ vs. data from the H1 collaboration \cite{H1:2005dtp}.
    Results for the $t$-integrated differential cross section in the photoproduction of
    $\psip(2S)$ are shown in the lower left panel and compared with data
    from the LHCb \cite{LHCb:2018rcm} collaboration.
    The lower right panel shows the model calculations of $d\sigma^{\gamma Pb\to \Jpsi Pb}(t)/dt$
    in comparison with data from the ALICE experiment \cite{Acharya:2021bnz}. 
  }
\EF

%
%
%
\section{Predictions within the Green function formalism}
\label{sec-data}
%
%
%

\subsection{Model calculations vs. data}
\label{data}

In the present paper we apply the recent GBS model 
with DGLAP evolution 
\cite{Golec-Biernat:2017lfv}
for the dipole-proton partial amplitude including correlation 
between the dipole transverse orientation $\vec r$ and 
the impact parameter of the collision $\vec b$, Eq.~(\ref{dipa-gbw}). 
Our calculations rely on the quarkonium wave functions generated by the 
POW $Q$-$\bar Q$ interaction potential 
\cite{Barik:1980ai}, 
since in combination with the GBS model provides the best description 
of the UPC data on rapidity distributions $d\sigma/dy$ (see Ref.~
\cite{Nemchik:2024lny}).
%
We also include the leading-twist gluon shadowing 
representing the main nuclear effect in the kinematic region 
when $x\lsim x_g$, where $x_g$ has the following form 
\cite{Ivanov:2002kc},
%
\BE
x_g
\approx
\frac{\sqrt{3}}{f_g\,m_N\,R_A^{ch}}\,.
  \label{rg}
\EE
%
Here, the factor $f_g = \la l_c\ra\,/\,\la l_c^G\ra \sim 10$ 
\cite{Kopeliovich:2000ra}
and $R_A^{ch}$ is the mean value of the nuclear charge radius.

As a first step, we verified in Fig.~\ref{Fig1-proton}
that the model calculations of $t$-distributions
$d\sigma^{\gamma^\ast p\to \Jpsi p}(t,W,Q^2\approx 0)/dt$, 
$d\sigma^{\gamma^\ast p\to \Jpsi p}(t,W,Q^2 = 8.9\,\GeV^2)/dt$,
as well as $t$-integrated
$\sigma^{\gamma^\ast p\to \psip p}(W,Q^2\approx 0)$ 
are in good agreement with the available data on the proton target
from the H1 
\cite{H1:2005dtp} 
and LHCb 
\cite{LHCb:2018rcm}
collaborations.
Then we checked that our calculations describe well the first ALICE data 
\cite{Acharya:2021bnz} 
on the $t$ dependence of coherent $\Jpsi$ photonuclear production,
as demonstrated in the lower right panel of Fig.~\ref{Fig1-proton}.
In the next sections, we focus only on the heavy gold target due to the expected 
maximum strength of all relevant nuclear phenomena, such as 
the CT and (reduced) CL effects, as well as the gluon shadowing.

\subsection{Manifestations of CT effects}
\label{CT}

\BF
%
\includegraphics[height=7.5cm]{Trt_psi1S_Q2_eik.eps}
\hspace*{0.10cm}
\includegraphics[height=7.5cm]{Trt_psi2S_Q2_eik.eps}
\\
\Caption{
\label{Fig2-ct}
    The $t$-dependent nuclear transparency, Eq.~(\ref{tr-t}), for the coherent electroproduction 
    of $\Jpsi(1S)$ (left panel) and $\psip(2S)$ (right panel)
    in the limit $l_c\gg R_A$ and at two fixed values of $Q^2 =0$
    and $50\,\GeV^2$.
  }
\EF

The effect of CT leads to a modification of the diffractive pattern in  
$d\sigma^{\gamma^{\ast} A\to V A}/dt$ which results in a shifted 
positions of the diffractive minima to larger values of $t$, as shown in Fig.~\ref{Fig2-ct}.
Here we present model calculations in the LCL regime ($l_c\gg R_A$) 
of the $t$-dependent nuclear transparency, defined as,
%
\BE
Tr_A^{coh}(t)
=
\frac
{d\sigma^{\gamma^{\ast} A\to V A}/dt}
{A^2\,\,d\sigma^{\gamma^{\ast} N\to V N}/dt\bigr |_{t=0}}
=
\frac
{\Bigl |\mathcal{A}^{\gamma^{\ast} A\to V A}(x,Q^2,t)\Bigr |^2}
{A^2 \Bigl |\mathcal{A}^{\gamma^{\ast} p\to V p}(x,Q^2,t=0)\Bigr |^2}
\,,
  \label{tr-t}
\EE
%
at two fixed values of $Q^2 = 0$ and $50\,\GeV^2$.
In the LCL regime given by Eqs.~(\ref{amp-A0m}) and (\ref{eik}), the amplitude
in the numerator of Eq.~(\ref{tr-t}) can be expressed as,
%
\BE
\mathcal{A}^{\gamma^{\ast} A\to V A}(x,Q^2,t = -q^2)
= 
2\,
\int d^2b_A\,e^{i\vec q\cdot\vec b_A}
\mathcal{M}_A(\vec b_A)\,.
  \label{bA}
\EE
%
To better understand the manifestation of CT effects, we can use
the quadratic approximation of the dipole cross section in Eq.~(\ref{eik}),
$\int d^2 b\,\mathrm{Im}\mathcal{A}_{Q\bar Q}^N(\vec r,x,\alpha,\vec b) \approx C(x) r^2/2$,
as well as the Gaussian approximation of the product of the photon and vector meson wave functions, $\Psi_{V}^{*}(\vec r,\alpha)\,\Psi_{\gamma^\ast}(\vec r,\alpha,Q^2) \propto \exp[ - r^2/\la r^2\ra]$.
Then the $b_A$- dependent amplitude in Eq.~(\ref{bA}) reads,
%
\BE
\mathcal{M}_A(\vec b_A)
=
\pi \la r^2\ra \cdot
\frac{C(x)\, T_A(\vec b_A)\,\la r^2\ra}
{2 + C(x)\, T_A(\vec b_A)\,\la r^2\ra}
\qquad
\text{and}
\qquad
\mathcal{A}^{\gamma^{\ast} p\to V p}(x,Q^2,t=0)
=
\pi\,C(x)\,\la r^2\ra^2
\,,
  \label{simp-A}
\EE
%
which gives the following simplified form for the $t$-dependent nuclear transparency, 
%
\BE
Tr_A^{coh}(t)
=
\frac
{4}{A^2} 
\biggl |
\int d^2b_A\,e^{i\vec q\cdot\vec b_A}
\frac{T_A(\vec b_A)}
{2 + C(x)\, T_A(\vec b_A)\,\la r^2\ra}
\biggr |^2\,.
  \label{tr-t-simp}
\EE
%

It can be seen from Eqs.~(\ref{simp-A}) and (\ref{tr-t-simp}) that
the partial amplitude $\mathcal{M}_A(\vec b_A)$ is suppressed with $Q^2$ due to diminishing
of the mean $Q\bar Q$-dipole size $\la r^2\ra$ and 
that the suppression is larger at smaller $b_A$-values.
Consequently, the slope $B_{\gamma^{\ast} A} = \la b^2\ra / 2$ 
of the differential cross section $d\sigma^{\gamma^{\ast} A\to V A}/dt$
should be reduced with $Q^2$, shifting the diffractive minima to
larger values of $t$, as indeed shown in Fig.~\ref{Fig2-ct}.
However, observing experimentally such a CT signal by $Q^2$ modification
of the diffractive pattern is questionable due to a rather weak effect. 
%
Here, the coherent production of light vector mesons gives a better chance to investigate 
the CT effects since their manifestations are more visible compared to the heavy quarkonium 
production, as analyzed in Refs.~
\cite{Kopeliovich:2001xj,Nemchik:2002ug}.
%
On the other hand, the CT effects could be recognized and measured 
by EIC experiments at RHIC and the LHeC as an increase in $t$-dependent
nuclear transparency $Tr_A^{coh}(t)$ with $Q^2$ at small fixed values of $t$, as predicted in the left panel of Fig.~\ref{Fig2-ct} for electroproduction of $\Jpsi(1S)$.
The coherent electroproduction of $\psip(2S)$ gives almost negligible 
chance for investigation of such effects (see the right panel of Fig.~\ref{Fig2-ct}).

\BF
%
\includegraphics[height=5.0cm]{Trt_psi1S_W-Q2_0-tfix.eps}
\hspace*{0.30cm}
\includegraphics[height=5.0cm]{Trt_psi2S_W-Q2_0-tfix.eps}
\hspace*{0.3cm}
\includegraphics[height=5.0cm]{Ratio-Q2-0.eps}
\\
\vspace*{0.6cm}
\includegraphics[height=5.0cm]{Trt_psi1S_W-Q2_50-tfix.eps}
\hspace*{0.30cm}
\includegraphics[height=5.0cm]{Trt_psi2S_W-Q2_50-tfix.eps}
\hspace*{0.3cm}
\includegraphics[height=5.0cm]{Ratio-Q2-50.eps}
\\
\vspace*{-0.0cm}
\Caption{
\label{Fig3-lc-reduced}
    Model predictions for the energy behavior of the $t$-dependent 
    differential cross section $d\sigma/dt$ for coherent photoproduction of 
    $\Jpsi(1S)$ (upper left panel), $\psip(2S)$ (upper middle panel)
    and for the $\psip(2S)$-to-$\Jpsi(1S)$ ratio of $d\sigma/dt$ (upper right panel) 
    at several fixed values of $t=0.001, 0.004, 0.008$ and $0.012\,\GeV^2$.
    The lower panels show the electroproduction process at $Q^2 = 50\,\GeV^2$.
    In all panels the Green function formalism with gluon shadowing corrections 
    (solid lines) is compared with the standard eikonal approximation without 
    gluon shadowing (dashed lines).
  }
\EF

\subsection{Reduced higher-twist corrections at EIC}
\label{quark-shadowing}

In the EIC kinematic region at RHIC corresponding to Bjorken 
$x\gsim x_c\approx 0.01$, where $x$ is given by Eq.~(\ref{x}), 
one should also include reduced quantum coherence effects for 
the lowest $|Q\bar Q\ra$ Fock state of the photon, i.e. going 
beyond the "frozen" eikonal approximation, Eq.~(\ref{eik}).
Instead of a standard approximate incorporation of such  effects
via the nuclear form factor depending on the longitudinal momentum 
transfer $q_L = 1/l_c$ 
\cite{Kopeliovich:1993gk}, 
we use a rigorous Green function formalism 
as described in the previous Sec.~\ref{dipole-A}. 
Such a formalism contains all multiple scattering terms and 
treats the nuclear form factor correctly.
The reduced quark shadowing significantly modifies the $t$-dependent  
charmonium yields compared to the eikonal approximation at small $W$ values, as can be seen in Fig.~\ref{Fig3-lc-reduced}. 
Here we present $d\sigma^{\gamma^*Au\to V Au}/dt$ for the coherent 
photoproduction of 1S and 2S charmonium states, 
$V = \Jpsi(1S)$ and $\psip(2S)$, on the gold target and 
their 2S-to-1S ratio as a function of the photon 
energy $W$ for several fixed values of $t$ (upper panels).
The lower panels of Fig.~\ref{Fig3-lc-reduced}
show the analogous model predictions as the upper panels but 
for the coherent electroproduction process
at $Q^2=50\,\GeV^2$.

Fig.~\ref{Fig3-lc-reduced} demonstrates that in the EIC kinematic 
region, the onset of reduced $l_c$-effects at the same photon 
energy is stronger in charmonium electroproduction compared 
to photoproduction process, in consistency with Eq.~(\ref{lc-qq}).
Whereas in the photoproduction limit ($Q^2\approx 0$) such effects 
can be experimentally recognized only at rather small $W\lsim 20\,\GeV$, 
i.e. at the lower limit of the kinematic region of future EIC experiments
(see the upper left and middle panels of Fig.~\ref{Fig3-lc-reduced}),
charmonium electroproduction at $Q^2 = 50\,\GeV^2$  provides 
a better chance for experimental investigation of reduced quark shadowing,
which occurs over a wider energy range, $W\lsim 50\,\GeV$  
(see differences between solid and dashed lines 
in the lower left and lower middle panels of Fig.~\ref{Fig3-lc-reduced}).

\subsection{
Searching for the gluon saturation at LHeC}
\label{gluon-shadowing}

The LHeC kinematic region provides a good basis for searching 
for possible manifestations of gluon saturation effects.
Here the charmonium photoproduction is more suitable compared to 
electroproduction process due to a stronger onset of the leading-twist shadowing corrections.
The corresponding factor $R_G$ in Eq.~(\ref{eq:dipole:gs:replace-b}) 
causes a sizable reduction of the gluon distribution function 
in a free nucleon at small dipole separations $r\lsim R_0(x,\mu^2)$ 
and only very slowly (logarithmically) is approaching unity as $Q^2\to\infty$.
The interpretation of the gluon shadowing phenomenon
is reference-frame dependent. 
It can be treated as glue-glue fusion in the infinite momentum 
frame of the nucleus.
The corresponding Lorentz contraction of the nucleus keeps 
still a sufficiently large separations of the bound nucleons, 
except the gluon clouds of the nucleons that are contracted 
less due to their smaller momentum fraction $x$ of the nucleon. 
Then gluons originating from different nucleons can overlap and interact with each other with their subsequent fusion
leading to a reduction in gluon density.
This corresponds to a non-linear QCD evolution incorporated 
into evolution equations
\cite{Gribov:1981ac,Gribov:1983ivg,balitsky,kovchegov}.

The leading-twist shadowing corrections in coherent charmonium 
photoproduction off nuclei, studied in the present paper, 
may represent an effective tool in searching for gluon saturation effects. 
Fig.~\ref{Fig3-lc-reduced} shows how the gluon shadowing phenomenon
affects the energy behavior of the $t$-dependent differential cross sections
(see the differences between solid and dashed lines at large photon energies)
in the production of 1S and 2S charmonium states and their 2S-to-1S ratios 
at different fixed values of $t$ and at $Q^2\approx 0$ (upper panels) and 
$Q^2 = 50\,\GeV^2$ (lower panels).
One can see that, within the LHeC energy range of $W\lsim 2000\,\GeV$, a monotonic rise of $d\sigma/dt$ with the photon energy at $t= 0$
is gradually changed for a non-monotonic shape with a maximum that 
is more visible at larger $t$ values and simultaneously shifted towards smaller values of $W$.
This result is consistent with an increase of the gluon
shadowing correction with $t$, since the $R_G(x,\vec b_A)$ factor
is scanned at smaller impact-parameter values $b_A$ (see Fig.~1 in Ref.~
\cite{Kopeliovich:2022jwe}).

The upper panels of Fig.~\ref{Fig3-lc-reduced} demonstrate that
in the photoproduction of charmonia off gold target the $t$-dependent 
positions of maxima in $d\sigma/dt(W)$ at $W=W_{max}$ cover a broad 
photon energy range, namely 
$W_{max}\sim 1500\,\GeV$ at $t = 0.004\,\GeV^2$,
$W_{max}\sim 800\,\GeV$ at $t = 0.008\,\GeV^2$ and
$W_{max}\sim 200\,\GeV$ at $t = 0.012\,\GeV^2$.
This gives an opportunity for a recognition of gluon saturation effects 
by future experiments at the LHeC.
However, finding such effects may require larger $t$ values and, consequently, the data with higher statistics.

Here we would like to note that the $b_A$ dependence of gluon shadowing, 
calculated in the present paper, is crucial for evaluation of 
the $t$-dependent differential cross sections $d\sigma/dt$. 
This is in contrast to the global data analyses providing only the 
$b_A$-integrated gluon shadowing.
Note that various models for the dipole-nucleon scattering amplitude can also slightly modify the kinematic region for the onset 
of saturation effects when a non-monotonic scenario for the energy 
dependence of $d\sigma/dt$ can be shifted to smaller or larger values of $t$.
The coherent charmonium electroproduction off nuclei reduces 
substantially the chance to investigate saturation effects compared 
to photoproduction reaction due to weaker leading-twist corrections 
corresponding to a weaker onset of the non-linear QCD evolution,  
as seen in lower panels of Fig.~\ref{Fig3-lc-reduced}.

The potential existence and position of the maxima in the
$W$ dependence of $d\sigma/dt$, as a manifestation of possible 
gluon saturation effects, depend on the magnitude of gluon shadowing.
Its determination is related to the distance $r_g$ of the gluon 
propagation from the $Q\bar Q$ pair, i.e. to the size of the $GG$ dipole.
The value of $r_g$ limits the validity of the approximation 
$Q\bar QG\sim GG$ used for calculation of gluon shadowing.
The corresponding condition reads $\eta^2\gg 1/r_g^2$, which
is safely satisfied in coherent charmonium photo- and 
electroproduction studied in this paper. 
In the opposite case the $Q\bar Q$ pair cannot be treated 
as a point-like object compared to the size of the entire 
$Q\bar QG$ Fock state of the photon.
The single diffractive data in hadronic collisions allow 
to extract a magnitude of $r_g$ as was done in Ref.~
\cite{Kopeliovich:1999am}.
%
Here the corresponding diffraction cross section $\propto r^4$ 
exhibits a larger sensitivity to the dipole separation 
$r$ compared to total cross section $\propto r^2$.
This led to the determination of the mean $GG$ dipole size 
of the order of $r_g\sim 0.3\,\fm$ 
\cite{Kopeliovich:1999am}.
%
Such a small gluon propagation radius represents 
so far the only way how to resolve the long-standing problem 
related to the small size of the triple-Pomeron coupling. 
The above value of $r_g$ is incorporated into the LF color 
dipole formalism via a nonperturbative $G$-$G$ interaction
using the path integral technique.

Note that the small value of $r_g$ is consistent with the results 
of other approaches, based on the instanton liquid model 
\cite{Schafer:1996wv},
on the QCD sum rule analysis of the gluonic formfactor of the proton 
\cite{Braun:1992jp},
as well as on the lattice calculations 
\cite{DElia:1997sdk}.
%
This indicates rather small leading-twist shadowing effects 
in accordance with the Next-to-Leading-Order analysis 
of available DIS data on nuclei 
\cite{florian}.

We predict a stronger energy dependence for 
$d\sigma^{\gamma Au\to \psip(2S) Au}/dt$ than for 
the coherent photoproduction of $\Jpsi(1S)$ due to 
the nodal structure of the radial wave function 
for the 2S charmonium state (compare the upper middle 
and upper left panel of Fig.~\ref{Fig3-lc-reduced}).
This also leads to an increase of the
$\psip(2S)$-to-$\Jpsi(1S)$ ratio $R_{\psip/\Jpsi}(t)$ of $t$-dependent differential 
cross sections with the photon energy, as demonstrated in the upper 
right panel of Fig.~\ref{Fig3-lc-reduced}.
The node effect for the 2S charmonium state gradually
fades with energy and we predict a similar scenario for a non-monotonic
energy behavior of $d\sigma/dt$ as for the 1S charmonium coherent
photoproduction (compare upper left and upper middle panels
of Fig.~\ref{Fig3-lc-reduced} at large $W$ values).
The corresponding maxima of $d\sigma/dt$ are shifted 
towards larger values of $W$,
$W_{max}\sim 1300\,\GeV$ at $t = 0.008\,\GeV^2$ and
$W_{max}\sim 500\,\GeV$ at $t = 0.012\,\GeV^2$.
Since the leading-twist shadowing corrections are very similar
for both the 1S and 2S charmonium states, the rise of
the ratio $R_{\psip/\Jpsi}(t)$ with energy does not differ much from that
without such non-linear QCD effects, as demonstrated in right panels 
of Fig.~\ref{Fig3-lc-reduced}.
The non-monotonic $W$-dependence of $d\sigma/dt$ for both the
1S and 2S chamonium states with different positions of maxima 
may generate non-monotonic energy behavior also
for the $R_{\psip/\Jpsi}(t)$, especially at larger $t$ values, as demonstrated in the upper right panel of
Fig.~\ref{Fig3-lc-reduced}.
However, it is questionable whether such a manifestation
of gluon saturation effects can be recognized experimentally
at the LHeC.

At large $Q^2=50\,\GeV^2$ the node effect for the $\psip(2S)$ 
state is very weak and we predict very similar scenarios for 
the $W$-dependent $d\sigma/dt$ in coherent electronuclear 
production of the both charmonium states, $\Jpsi(1S)$ and $\psip(2S)$ 
(see the lower left and lower middle panel of Fig.~\ref{Fig3-lc-reduced}).
This also leads to a very weak energy dependence of the ratio $R_{\psip/\Jpsi}(t)$
(see the lower right panel of Fig.~\ref{Fig3-lc-reduced}).
Here we present the results only for fixed $t=0.001$ and $t=0.012\,\GeV^2$ 
because the curves at all values of $t$ lie very close to each other.

Note that in all calculations of differential cross
sections on nuclear targets, we have verified that in
comparison with the proton target the
effect of the correlation between the dipole
orientation and impact parameter of a collision
is minimal, causing a very small reduction 
of $d\sigma^{\gamma^{\ast} Au\to \Jpsi(1S) Au}/dt$ and
$d\sigma^{\gamma^{\ast} Au\to \psip(2S) Au}/dt$ by only a few percents.

In the present work, we tried to minimize the theoretical uncertainties
related to the elimination of D-wave admixture in charmonium wave functions,
to the choice of realistic $Q-\bar Q$ potential together with the Melosh
spin effects, to the inclusion of dipole orientation with respect 
to the impact parameter of the collision, 
as well as to the proper calculations of the leading-twist shadowing 
corrections consistent with other models and available diffractive data.
We believe that our predictions for manifestation of gluon saturation effects
are realistic and can be verified  by the future LHeC experiments.

\subsection{Pitfalls in the search for gluon saturation effects}
\label{pitfalls}

Here we briefly mention and comment on several aspects that may 
affect the shape of the $t$-dependent and $t$-integrated 
cross sections on the proton and nuclear targets
in the kinematic region of small $x\lsim 10^{-4}\div 10^{-5}$. 
As an example we refer to the recent work
\cite{Cepila:2025rkn} 
(see also references therein) devoted to search
for gluon saturation effects. 
The detailed analysis of all pitfalls,
which are mentioned below, will be presented elsewhere.

\vspace*{0.3cm}
{\bf i) Shape of quarkonium wave functions}
\\

Investigation of heavy quarkonium electroproduction off protons 
and nuclei is a very effective tool in searching for manifestations 
of saturation effects, since the uncertainties in the corresponding 
theoretical description are reduced
\cite{Cepila:2019skb} 
in comparison with light vector mesons.
In addition, the node effect in radially excited heavy quarkonium states
gives rise to a sensitivity to the choice of the LF quarkonium wave functions.
In Ref.~
\cite{Cepila:2025rkn}, 
the predictions for a possible signal of gluon saturation effects in terms 
of the $R_{\psip/\Jpsi}(t)$ ratio are based on the Gaussian-like wave functions related to harmonic oscillatory (HO) $Q$-$\bar Q$ interaction potential.
Moreover, the Coulomb term, first introduced in the LF HO model 
for the vector meson wave functions more than thirty years ago 
in Refs.~
\cite{Nemchik:1994fp,Nemchik:1996cw},
has been ignored. 
However, a strong evidence of such Coulomb contribution may be important 
at small dipole separations related to electroproduction of charmonia 
at large $Q^2\gg M_{\Jpsi}^2$ and especially in the photoproduction of bottomonia, 
which is one of the subjects of study in Ref.~
\cite{Cepila:2025rkn}.
%
Moreover, the absence of the Coulomb correction can affect 
the node position in radial wave functions mainly for bottomonium 
excited states, such as $\Yp(2S)$ and $\Ypp(3S)$.
Consequently, this may have a large impact on manifestations of gluon saturation effects in the analysis of the energy dependence
of ratios $R_{\Yp/\Y}(t)$ and $R_{\Ypp/\Y}(t)$, as was done in Ref.~
\cite{Cepila:2025rkn}.

Besides, the HO wave functions are not fully appropriate also
for description of charmonium states, since they can lead, for example, 
to an unexpectedly strong enhancement (possible anti-shadowing effects) 
in the incoherent photoproduction  of $\psip(2S)$ on nuclear targets 
(see e.g., Ref.~
\cite{Kopeliovich:1991pu}).
%
Moreover, the HO model is not able to describe the data on the $\psip(2S)$-to-$\Jpsi(1S)$ 
ratio of electroproduction cross sections as a function of the photon energy and virtuality, as can be seen in Fig.~21 of Ref.~
\cite{Cepila:2019skb}.

Instead, several realistic potential models can be found in the literature 
\cite{Barik:1980ai,Eichten:1979ms,Eichten:1978tg,Quigg:1977dd,Buchmuller:1980su},
which include also the Coulomb correction and describe the available data 
well, as shown in Ref.~
\cite{Cepila:2019skb}.
%
Fig.~21 in Ref.~
\cite{Cepila:2019skb} 
also demonstrates a large deviation between predictions 
based on realistic and HO potential models.
Such observations rule out the universal application of the HO 
potential model, frequently used in the literature even without 
the Coulomb contribution, for description of heavy quarkonium 
electroproduction.
For this reason, the
signal of gluon saturation effects, found in model predictions of Ref.~
\cite{Cepila:2025rkn}
should be additionally verified by calculations using other models
for the $c$-$\bar c$ and $b$-$\bar b$ potentials.

\vspace*{0.3cm}
{\bf ii) D-wave admixture in charmonium wave functions}
\\

Another pitfall in the investigation of gluon saturation effects
concerns the unjustified D-wave admixture in charmonium wave functions, 
related to the photon-like structure of the $V\to Q\bar Q$ transition.
It leads to a large $20\div 30\%$ enhancement of the 
$\psip(2S)$-to-$\Jpsi(1S)$ ratio $R_{\psip/\Jpsi}(W)$ of production cross sections, 
as demonstrated in Ref.~
\cite{Krelina:2020bxt}.
%
Such a D-wave effect is comparable and/or even stronger than that 
responsible for a contribution of the non-linear term  
in the Balitsky-Kovchegov evolution equation 
\cite{balitsky,kovchegov}, 
which leads to a stronger increase of $R_{\psip/\Jpsi}(W)$ with energy $W$,
as analyzed in Refs.~
\cite{Peredo:2023oym,Cepila:2025rkn}.
%
Thus, a significant reduction of D-wave components 
in quarkonium wave functions, in consistency with
solutions of the Schr\"odinger equation with realistic $Q\bar Q$ potential
models in the rest frame
\cite{Haysak:2003yf,Fu:2018yxq,Chang:2010kj,Cao:2012du}, 
causes a decrease in the ratio $R_{\psip/\Jpsi}(W)$.
This may lead to a complete elimination of any signal 
related to gluon saturation effects.

\vspace*{0.3cm}
{\bf iii) $\vec r$-$\vec b$ correlation}
\\

The shape of $d\sigma^{\gamma^{\ast} p \to \Jpsi(1S) p}/dt$ 
is closely related to $\vec r$-$\vec b$ correlation which 
should represent an indispensable feature of the 
$\vec b$-dependent dipole-proton partial amplitude.
The corresponding form of such an amplitude should lead, 
after integration over the impact parameter of the collision $\vec b$, 
to a correct standard dipole-proton cross section of a saturated form 
at $r\gg R_0(x)$ with original parameters extracted from 
deep-inelastic-scattering HERA data on the proton structure 
functions (see Sec.~\ref{dipole-p}). 
This is not the case of parametrization of 
$\mathrm{Im} \mathcal{A}_{\bar QQ}^{N}(\vec r, x_0=0.01, \vec{b}\,)$ 
presented in Ref.~
\cite{Cepila:2025rkn}
as the initial condition for the BK equation, denoted as the CCV model.
Here, performing integration, 
$2\, \int d^2 b\,\mathrm{Im} \mathcal{A}_{\bar QQ}^{N}(\vec r, x_0, \vec{b}\,)$, 
one obtains a pathological non-monotonic behavior of the dipole cross 
section $\sigma_{Q\bar Q}(r,x_0)$ as a function of the dipole size $r$,
when $\sigma_{Q\bar Q}(r,x_0)\to 0$ at large $r$
in contrast with the onset of saturation.
Moreover, the CCV model also disregards the dependence on the
fractional momenta $\alpha$ of $Q$ or $\bar Q$. 
These facts arise the question whether such $\vec r$-$\vec b$ 
correlation has been included properly.

To clarify this, in Fig.~\ref{Fig-br-comp} we compare the CCV
model with that used in the present paper, Eq.~(\ref{dipa-gbw}), 
denoted as the KKN model.
\cite{Kopeliovich:2008dy,Kopeliovich:2007sd,Kopeliovich:2007fv,
Kopeliovich:2008nx,Kopeliovich:2021dgx}. 
%
Here we present the ratio $R_{\perp/\parallel}(b) = 
\mathrm{Im} A_{Q\bar Q}^N(\vec r,\vec b,\Theta=\pi/2) / 
\mathrm{Im} A_{Q\bar Q}^N(\vec r,\vec b,\Theta=0)\equiv 
N(b,\Theta=\pi/2)/N(b,\Theta=0)$ 
as a function of the impact parameter of a collision $b$ 
at fixed values of the dipole size corresponding to the scanning radius $r_S$ 
\cite{Nemchik:1994fp} 
in production of $\rho^0$ (dotted lines), $\Jpsi$ (dashed lines) 
and $\Y$ (solid lines) mesons, where $r_S = Y/\sqrt{M_V^2 + Q^2}$ 
with $Y_{\rho} = 7$, $Y_{\Jpsi}=6$ and $Y_{\Y}=6$.

\BF
%
\includegraphics[height=7.5cm]{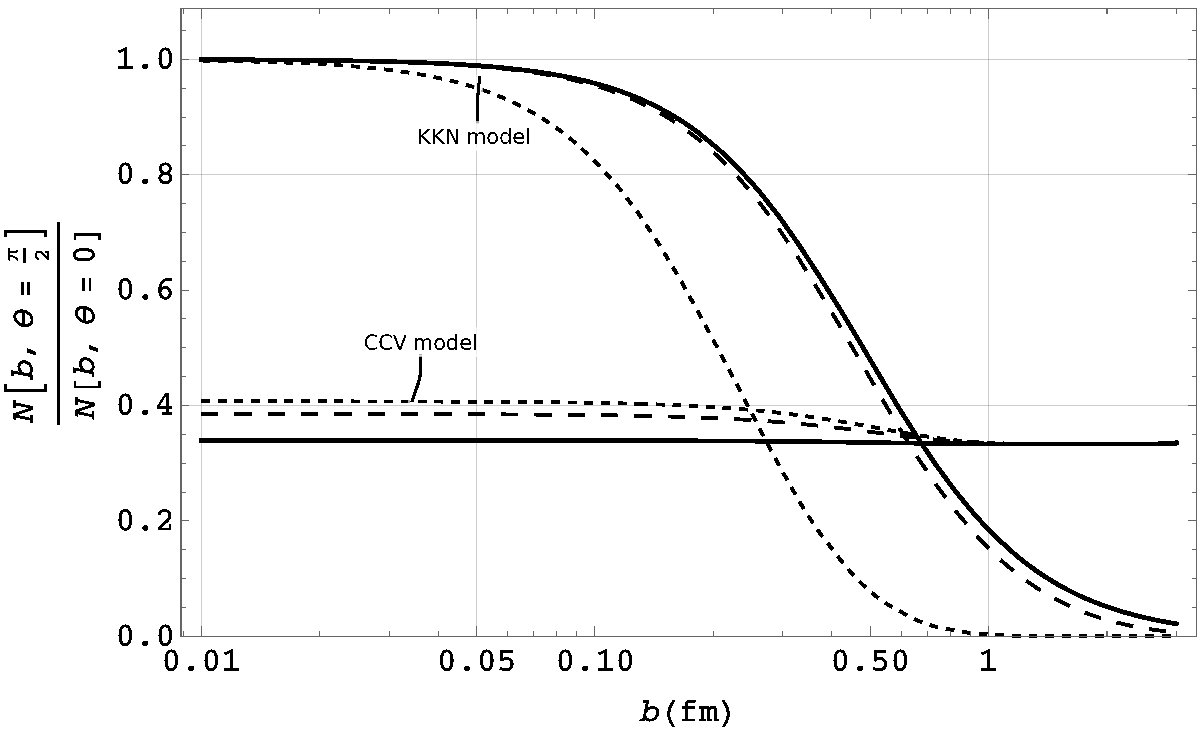}
\Caption{
  \label{Fig-br-comp}
(Color online) 
          The comparison of the CCV and KKN model predictions for the ratio 
          $R_{\perp/\parallel}(b)$ as a function of the impact parameter of a collision $b$. 
          Here the solid, dashed, and dotted lines correspond to the production of 
          $\Y$, $\Jpsi$, and $\rho^0$ mesons, respectively.
}
\EF

The KKN model in Fig.~\ref{Fig-br-comp} correctly reproduces the
limits $b\to 0$ and $b\gg R_0(x_0)$. 
In the case when $b\to 0$, one cannot expect
any vanishing interaction, when $\vec r\perp\vec b$. 
Here the $\vec r$-$\vec b$
correlation becomes irrelevant and the ratio $R_{\perp/\parallel}(b\to0)\to 1$.
However, with increasing $b$, the orientation $\vec r\perp\vec b$
gradually comes into play causing a decreasing $R_{\perp/\parallel}(b)$,
which finally tends to zero at very large $b$-values, when the 
$Q\bar Q$ dipole-nucleon interaction with $\vec r\perp\vec b$ 
completely ceases compared to $\vec r\parallel\vec b$ orientation. 
Fig.~\ref{Fig-br-comp} also demonstrates a correct hierarchy
in production of various vector mesons, namely that the ratio 
$R_{\perp/\parallel}(b)$  starts to decrease with $b$ firstly for $\rho^0$
mesons with the largest scanning radius and then for $\Jpsi$ and $\Y$ mesons.
In contrast to the KKN model, the CCV model is practically insensitive 
to $b$-values, leading to an inverse hierarchy of 
$R_{\perp/\parallel}(b)$ values in production of various quarkonia.
For this reason, we consider this CCV model for the  $\vec r$-$\vec b$ 
correlation as incorrect.

The recent studies 
\cite{Peredo:2023oym,Cepila:2025rkn}
propose to investigate the $\psip(2S)$-to-$\Jpsi(1S)$ ratio 
$R_{\psip/\Jpsi}(t,W)$ of the cross sections as a possible 
manifestation of gluon saturation effects.
The onset of such effects should lead to a much stronger rise of $R_{\psip/\Jpsi}(t,W)$ with the photon energy compared to 
the case where no saturation effects are involved. 
However, we have found in Refs.~
\cite{Kopeliovich:2021dgx,Krelina:2021jzn}
that $\vec r$-$\vec b$ correlation significantly modifies the 
shape of $t$-dependent differential cross sections, reducing 
$R_{\psip/\Jpsi}(t,W)$ and causing a more flat $t$ dependence and
a stronger rise with the photon energy compared to the case 
when vectors $\vec b$ and $\vec r$ are parallel.
Thus, a correct incorporation of $\vec r$-$\vec b$ correlation 
effects is unavoidable in calculations and may affect 
substantially the expected signal of gluon saturation.
Specific detailed calculations will be presented in our forthcoming paper.

\vspace*{0.3cm}
{\bf iv) Kernel in the BK equation}\\

The solution of the BK evolution equation depends strongly on the Kernel,
which has been adopted, e.g. in Ref.~
\cite{Cepila:2025rkn} 
in the form, which is inconsistent with the finite gluon propagation 
radius supported by data on gluon radiation 
\cite{Kopeliovich:1999am,Kopeliovich:2007pq}.
%
Calculations should be performed with BK Kernel corresponding 
to non-perturbative quark-gluon wave function 
\cite{Kopeliovich:1999am}
due to rather large mean quark-gluon separation $r_g\sim 0.3\,\fm$, 
which is independent of $m_Q$ and/or $Q^2$ (up to Log corrections), but is
dependent on the scale, the effective gluon mass, $m_g \approx 0.7\,\GeV$.
This may significantly modify the model predictions for 
$d\sigma^{\gamma^{\ast} p \to \Jpsi(1S) p}/dt$ 
compared to the results presented in Ref.~
\cite{Cepila:2025rkn}.
%
Such a modification may affect conclusions about manifestations 
of gluon saturation effects.

\vspace*{0.3cm}
{\bf v) Reduced shadowing effects in the BK equation 
        for the dipole-nucleus amplitude}\\

In description of heavy quarkonium electroproduction off nuclei, 
the BK equation for the eikonalized dipole-nucleus partial 
amplitude, Eq.~(\ref{eik}), cannot be applied to nuclear targets. 
The reason is that all dipoles with different numbers of
gluons are treated as long-lived photon fluctuations 
contributing to nuclear shadowing in a maximal strength. 
Such a result does not correspond to the fact that each additional 
gluon in the photon Fock state significantly reduces the corresponding coherence 
length, thus causing a gradual diminishing of contributions 
to shadowing effects from the photon components containing 
more and more gluons (see Sec. IV B in Ref.~
\cite{Kopeliovich:2022jwe}). 
%
As a consequence, the BK equation can significantly overestimate 
the nuclear shadowing at large $W$, which may have a direct impact on
the search for gluon saturation effects.
The first simplified calculations in Ref.~
\cite{Nemchik:2022kkx}
show that in the LHC kinematic region the BK equation in combination 
with the frequently used eikonal form, Eq.~(\ref{eik}), for the 
dipole-nucleus partial amplitude leads to an overestimation 
of the shadowing effects by about 20$\,\%$.
The modification of the BK evolution equation to include such reduced 
coherence effects will be presented elsewhere.

%
%
%
%
%
\section{Summary}
\label{sec-sum}
\vspace*{-0.2cm}
%
%
%
%
%

The transverse momentum transfer dependence of differential 
cross sections $d\sigma/dt$ for the coherent electroproduction 
of heavy quarkonia on nuclei was studied in the framework of the dipole 
description based on a rigorous Green function formalism.
We have analyzed a significance of the reduced effects of quantum 
coherence in the EIC kinematic region at RHIC, as well as leading-twist 
corrections at LHeC energies. The latter effect can also be treated
as a manifestation of non-linear QCD effects related to gluon saturation.

In order to minimize theoretical uncertainties in our predictions,
we consider the following:

\begin{itemize}
\vspace*{-0.3cm}
\item
We include the correlation between impact parameter of a collision 
$\vec b$ and the dipole orientation $\vec r$.
\vspace*{-0.25cm}
\item
We use the structure for the $V\to Q\bar Q$ transition
which eliminates the exaggerated weight of the D-wave component 
in the rest frame quarkonium wave function, in consistency with
the solutions of the Schr\'odinger equation.
\vspace*{-0.25cm}
\item
We rely on LF wave functions of heavy quarkonia generated 
from a solution of the Lorentz boosted Schr\"odinger equation with 
realistic potentials together with Melosh spin rotation effects.
\end{itemize}
\vspace*{-0.25cm}

Each Fock component of the incoming photon, 
i.e. $|\bar QQ\ra$, $|\bar QQG\ra$, $|\bar QQ2G\ra$, ..., 
contributes independently to the heavy quarkonium electroproduction
on nuclear targets in accordance with the corresponding coherence lengths. 
The UPC energy range at the LHC allows the eikonalization 
of the partial $\vec b$-dependent dipole $Q \bar Q$-proton amplitude 
(see Eq.~(\ref{eik})), since the coherence length for the lowest $Q\bar Q$ 
Fock state considerably exceeds the nuclear size, $l_c\gg R_A$. 
The corresponding quark shadowing gradually diminishes with the 
heavy quarkonium mass and thus represents a higher-twist effect.

However, a part of the kinematic region of the prepared EIC 
experiments at RHIC will cover quite large $x\gsim~x_c\approx 0.01$, 
and one eventually has to deal with reduced coherence effects, 
when $l_c\lsim  R_A$. 
Therefore, we apply a rigorous path integral technique for calculations 
of $t$-dependent differential cross sections and derive the corresponding 
formulas (see Eqs.~(\ref{amp-A-sr1}) and (\ref{amp-A-sr2})).
The incorporation of the reduced $l_c$ effects represents 
the main advancement of our present work. 
The reduced $l_c$ effects lead to a significant decrease in the yields 
of $t$-dependent charmonium production.
The corresponding predictions for $d\sigma^{\gamma^{\ast} A\to V A}/dt$ 
($V = \Jpsi(1S), \psip(2S)$) can be tested by future EIC experiments at RHIC.

The high-energy part of the EIC kinematic region, when
$x\lsim x_g$, requires also to include shadowing 
corrections from the higher Fock components of the photon, 
$|\bar QQG\ra$, $|\bar QQ2G\ra$, etc.
Such corrections are known as the gluon shadowing, which
is the leading-twist effect because the transverse size 
of the $Q\bar Q$-$G$ dipoles is large and nearly scale independent.
They are much stronger than the higher-twist shadowing
and their calculations have to rely on the 
path integral technique, because the dipoles $Q\bar Q-G, ...,$ 
cannot be treated as "frozen" even at very high energies 
due to a divergent $d\alpha_g/\alpha_g$ behavior.

Besides effects of quantum coherence, 
the Green function formalism also allows us to study the manifestations
of color transparency effects as a modification of the shape
of $d\sigma/dt$ and/or $Tr_A^{coh}(t)$ with $Q^2$. 
We have found that such effects lead to a shift of the diffraction 
minima towards larger values of $t$ (see Fig.~\ref{Fig2-ct}), 
caused by a $Q^2$-reduction of the slope $B_{\gamma^{\ast}A}(Q^2)$ 
of the $t$-distributions.
However, it will be difficult to detect it experimentally. 
The only way to detect CT effects by EIC experiments at RHIC 
and the LHeC is to measure an increase of $t$-dependent nuclear 
transparency with $Q^2$ at fixed values of $t$, as seen in Fig.~\ref{Fig2-ct} 
for the electroproduction of $\Jpsi(1S)$.

Coherent photoproduction of charmonia off nuclei at large LHeC energies, 
where the coherence length $l_c^G\gg 2\,\fm$,
favors to analyze the dominant role of gluon shadowing corrections 
allowing a subsequent study of a significance of gluon saturation effects.
We predict a non-monotonic energy dependence of $d\sigma/dt$
(see Fig.~\ref{Fig3-lc-reduced}) that is more visible at larger $t$ values 
with corresponding maxima shifted towards smaller photon energies.
Such an energy behavior of $d\sigma/dt$ can be verified by future 
EIC experiments at planned LHeC.
This can shed more light on the magnitude of the leading twist
gluon shadowing and can help us to rule out various 
phenomenological models devoted to study of nuclear shadowing.
Specifically, this will also contribute to resolving the issue 
of the applicability of the BK evolution equation in its current 
form in combination with the high-energy eikonal  
approximation for the partial dipole-nucleus amplitude for description
of processes on nuclear targets, frequently used in the literature.

We briefly discussed that conclusions about the manifestations 
of possible gluon saturation effects should be taken with a grain of salt 
until the pitfalls that cause significant modification 
of quarkonium production rates are resolved and clarified,
such as
the shape of the $Q$-$\bar Q$ interaction potential, 
an unjustified enhanced D-wave admixture in quarkonium wave functions, 
a correct model for the correlation between dipole orientation and 
impact parameters of a collision, 
the form of the realistic Kernel in the BK equation
consistent with the finite gluon propagation radius, 
and reduced effects of quantum coherence in the BK equation.
The study of radially excited heavy quarkonia at the EIC will 
provide a greater sensitivity to the manifestation of all these 
particular phenomena compared to production of 1S ground states.   
This can represent a very powerful tool for ruling out various 
phenomenological models that are seeking conclusive evidence 
of saturation effects.

%
%
%
\begin{acknowledgments}
\vspace*{-0.2cm}
The work of J.N. was partially supported 
by the Slovak Funding Agency, Grant No. 2/0020/22.
Computational resources were provided by the e-INFRA CZ project (ID:90254), supported by the Ministry of Education, Youth and Sports of the Czech Republic.
\end{acknowledgments}
%
%
%

%
%
%
%
%


\end{document}